# Occurrence of Great Magnetic Storms on 6–8 March 1582

Kentaro Hattori[1], Hisashi Hayakawa[2,3] and Yusuke Ebihara[4]

[1]*Kyoto University, Department of Geology and Mineralogy, Graduate School of Science, Japan*

[2]*Osaka University, Graduate School of Letters, Toyonaka, Osaka, Japan*

[3]*Rutherford Appleton Laboratory, Science and Technology Facilities Council, RAL Space Didcot, Oxfordshire, UK*

[4]*Kyoto University, Research Institute for Survival Hemisphere Kyoto, Japan*

**ABSTRACT**

Although knowing the occurrence frequency of severe space weather events is important for a modern society, it is insufficiently known due to the lack of magnetic or sunspot observations, before the Carrington event in 1859 known as one of the largest events during the last two centuries. Here, we show that a severe magnetic storm occurred on 8 March 1582 based on auroral records in East Asia. The equatorward boundary of auroral visibility reached 28.8° magnetic latitude. The equatorward boundary of the auroral oval is estimated to be 33.0° invariant latitude (ILAT), which is comparable to the storms on 25/26 September 1909 (~31.6° ILAT, minimum Dst of −595 nT), 28/29 August 1859 (~36.5° ILAT), and 13/14 March 1989 (~40° ILAT, minimum Dst of −589 nT). Assuming that the equatorward boundary is a proxy for the scale of magnetic storms, we presume that the storm on March 1582 was severe. We also found that the storm on March 1582 lasted, at least, for three days by combining European records. The auroral oval stayed at mid-latitude for the first two days and moved to low-latitude (in East Asia) for the last day. It is plausible that the storm was caused by a series of ICMEs (interplanetary coronal mass ejections). We can reasonably speculate that a first ICME could have cleaned up interplanetary space to make the following ICMEs more geo-effective, as probably occurred in the Carrington and Halloween storms.

**Key words:** Sun: activity – Sun: coronal mass ejections (CMEs) – Sun: flares – solar-terrestrial relations – planets and satellites: aurorae – planets and satellites: magnetic fields

**1 INTRODUCTION**

Studies on extreme space weather events are of significant importance for modern society, which is increasingly dependent on technological infrastructures (Daglis 2000, 2004; Baker et al. 2008; Hapgood 2011; Riley et al. 2018). To date, the magnetic storm on 13/14 March 1989 with a minimum Dst value (−589 nT) is the largest one since 1957, when the measurement of Dst index was started (Sugiura 1964; Sugiura & Kamei 1991). The storm on 13/14 March 1989 caused not only an auroral display at low magnetic latitudes (MLATs), but also a serious blackout in Québec (Allen et al. 1989; Bolduc 2002).

Arguably, the Carrington event in September 1859 is considered as one of the greatest magnetic storms in observational history, because of the extreme equatorward extension of its auroral oval and the magnetic disturbances at low latitudes (Tsurutani et al. 2003; Cliver & Svalgaard 2004; Green & Boardsen 2006; Silverman 2006; Siscoe, Crooker & Clauer 2006; Cliver & Dietrich 2013; Lakhina & Tsurutani 2016; Hayakawa et al. 2016a, 2018d). Indeed, the Carrington storm even caused a serious disturbance in the low technology telegraph systems of North America and Western Europe (Boteler 2006; Muller 2014).

Thus, recent studies warn us that the great magnetic storms, such as Carrington storm, would be catastrophic to the modern society (Baker et al. 2008; Hapgood 2011). A question arises of how often a Carrington-class storm occurs. Statistical studies show us that a Carrington event seems to be a once-a-century event in terms its geomagnetic disturbance and solar flare effect (Riley 2012; Riley & Love 2017; Curto, Castell & Del Moral 2016). However, historical evidence suggests that extreme events of similar scale have occurred even more frequently (Kappenman 2006; Cliver & Dietrich 2013; Silverman & Cliver 2001; Silverman 2008; Hayakawa et al. 2018b, 2018d).

Due to its importance and rarity, the temporal coverage of data is insufficient. Systematic magnetic observation starts only after the mid-19th century (Nevanlinna 2006; Araki 2014; Lockwood et al. 2018a, 2018b) and sometimes suffers off-scale measurements during extreme events (Boteler 2006; Love, Hayakawa & Cliver 2019; Hayakawa et al. 2019a). Telescopic sunspot observation is considered to be one of the longest ongoing experiments in human history, but measurements span only four centuries since 1610 (Vaquero & Vázquez 2009; Owens 2013; Clette et al. 2014; Vaquero et al. 2016).

To overcome this issue, the equatorward boundary of auroral oval can be used for a proxy measure of the strength of magnetic storms (cf. Yokoyama, Kamide & Miyaoka 1998). Auroral records and drawings let us trace back the history of magnetic storms to 771/772 and even 567 BC (Stephenson, Willis & Hallinan 2004; Hayakawa et al. 2016b, 2017b), and examine cases of space weather events in ancient to medieval epochs (Usoskin et al. 2013; Hayakawa et al. 2017a, 2017b, 2019b; Stephenson et al. 2019). Because an aurora is a global phenomenon, simultaneous auroral observations at different sites (different latitudes and longitudes) can be used to distinguish the aurora from other, local phenomena (Willis Stephenson & Singh 1996; Willis & Stephenson 2000; Hayakawa et al. 2017a, 2017d, 2018a; cf. Silverman 2003; Vaquero et al. 2008; Hayakawa et al. 2018c).

In this paper, we focused on the auroral records of March 1582. Willis & Stephenson (2000) identified simultaneous records in Japan and China for this period. We extended their study with new records and compared them with other records in Western Europe. By computing MLATs of the observing sites, we estimated the equatorward boundary of the auroral oval using information about colour and elevation angle of the aurorae.

**2 METHOD**

We surveyed contemporary historical documents in the early half of March 1582 in East Asia. We consulted the official histories, imperial records, and local treatises in China, imperial records and personal diaries in Korea, and diaries and chronicles in Japan (cf. Appendix 1 for a bibliography). We identified these observing sites and computed the MLAT at that time based on the contemporary location of the magnetic north pole (N85°25′, W34°30′) according to the archaeomagnetic field model Cals3k.4b (Korte & Constable 2011). We then compared the results with other European records in this epoch.

**3 RESULTS**

Willis & Stephenson (2000) used one Chinese report and two Japanese reports from 8 March 1582. Here, we surveyed two Chinese reports (Beijing Observatory 1988), two

Korean reports, and seven Japanese reports as shown in Table 1 and Figure 1. In this epoch, the number of the local reports had become larger than that of the official histories. The authorized historical documents consist of imperial chronicle, biographies of nobles and notables, treatises, tables, and so on (Willis & Stephenson 2000; Lee et al. 2004; Hayakawa et al. 2017c). We also surveyed 39 reports from Western Europe listed in Table 2. Their bibliographic details are given in Appendices 1 and 2 and their geographical locations are shown in Figures 2 and 3.

    We identified the observing sites according to the location of their authors at that time, unless the observing sites are explicitly clarified in the text. As to diaries, the described events were mostly those observed by the authors themselves, and hence associated with the place where the author lived at that time.

    Willis & Stephenson (2000) assumed that the two Japanese observations were made on Honshu Island, based on Kanda (1935). Because Honshu is ~1,500 km long, a more specific identification is required to estimate the equatorward boundary of the auroral oval. By surveying the original historical documents, we identified the specific locations of the observing sites of Figures 2 and 3.

## 4 EQUATORWARD EXTENSION OF AURORAL OVAL ON 8 MARCH 1582 IN EAST ASIA

As shown in Table 1, the observing sites of the aurora in East Asia on 8 March 1582 are distributed from 28.8° MLAT (J07, Búngo) to 35.6° MLAT (C01). To identify the equatorward boundary of the auroral oval, we need the MLAT of the observing site as well as the elevation angle of the aurora. Investigating the original description in detail, we found the following reports with information of elevation angle: 'there were many stripes in vermillion colour extending over half of the upper sky' (J05; N34°41′ E135°50′, 30.2° MLAT), 'they were reddish and bluish, and disappeared after arriving southward at the dawn' (K01; N36°41′ E128°25′, 32.3° MLAT), and 'red and white cloud vapour surrounded half of the heaven' (C01; N38°56′ E100°27′, 35.6° MLAT). Interpreting these records literally, we can reasonably consider that the aurorae extended equatorward at least beyond the zenith at each observing site.

    Based on the MLAT of observing sites and the elevation angle, we can estimate their equatorward boundaries in terms of invariant latitudes (ILATs) (cf. Figure 4). ILAT $\Lambda$ is used to label a magnetic field line (O'Brien et al. 1962), and is given by

$$\cos^2 \Lambda = \frac{1}{L}, \tag{1}$$

where $L$ is McIlwain's L-value (McIlwain 1961). MLAT $\lambda$ is given by

$$\cos^2 \lambda = \frac{R}{aL}, \tag{2}$$

where $R$ and $a$ are the radial distance from the centre of the Earth, and the radius of the Earth, respectively (cf. Hayakawa et al. 2018b, 2018c, 2019a). ILAT has the same value along the magnetic field line labelled by $L$. For a dipole magnetic field, ILAT is the same as MLAT on the ground (where $R = a$). From the descriptions in J05, K01 and C01, it is reasonably supposed that the aurora reached the zenith. From the equations (1) and (2), with MLAT at the observing site ($\lambda_0$), the upper limits of ILATs ($\Lambda$) of the aurorae seen at J05, K01 and C01 are calculated to be 33.0°, 34.9°, and 37.9°, respectively. Here, the magnetic field and the height of the aurorae ($h$) are assumed to be the dipole and 400 km, respectively (Silverman 1998; Ebihara et al. 2017). The different ILATs may be attributed to temporal and spatial variations of the auroral oval.

According to the Korean record (K01) and the Chinese record (C01), the aurora consists of reddish, bluish, and whitish colours. Therefore, these two records likely describe an aurora arising from precipitation of energetic particle rather than a stable red auroral (SAR) arc (cf. Kozyra, Nagy & Slater 1997; Shiokawa, Ogawa & Kamide 2005). The SAR arc is thought to be a response of the upper atmosphere to electron heat flows, or precipitation of electrons with very low energies (Cole 1965). When temperature of the thermal electrons is very low, they cannot excite the oxygen atoms to the $^1S$ state, which is necessary to emit photons at 557.7 nm (green line) because the number of electrons with energy $\geq 4.12$ eV is very low (Kozyra et al. 1997). The thermal electrons excite the oxygen atoms to the $^1D$ state efficiently, giving rise to the emission at 630.0 nm (red line) selectively. The presence of the colour other than red is used to distinguish auroral phenomena from the SAR arc (Shiokawa et al. 2005).

As for the Japanese record (J05), only the reddish colour is described. However, based on the ray structure described, it is believed the Japanese record also describes an aurora. This is because SAR arcs are less structured. Stripes with reddish colour are often observable at low MLATs during large magnetic storms, such as the storm on 17 September 1770 (cf. Figure 1 of Hayakawa et al. 2017d). Therefore, we consider that the equatorward boundary of the auroral oval can be reasonably estimated as 33.0° ILAT (J05).

**5 AURORAL OBSERVATIONS ON 6–8 MARCH 1582 IN EUROPE**

During the same night, auroral displays were also reported in Europe (Table 2). Willis & Stephenson (2000) cited the Prussian record from Fritz (1873) with a caveat on its bibliographical uncertainty. We carefully surveyed the original reference (E19) of this Prussian report and added seven records from Zürich (E05), Göttingen (E34), Provins (E35), Paris (E36), Azay-sur-Cher (E37), Anvers (E38) and London (E39). These auroral records indicate that the aurora was witnessed globally, in Europe and Eastern Asia, on 8 March 1582.

Previously, auroral displays were reported on 6 and 8 March as noted by Willis & Stephenson (2000) based on Link (1962) and Fritz (1873). According to the newly re-discovered reports (E21, E31–E33), the aurora was also visible on 7 March, meaning that the aurora was successively visible at least for 3 nights in Europe (Table 2). The equatorwardmost boundaries of the auroral visibility are reported as follows: Madrid (E29, E30; N40°25', W003°42', 44.3° MLAT) on 6 March, Cortona (E32; N43°17', E011°59', 46.3° MLAT) on 7 March, and Zürich (E05; N47°23', E008°33', 50.6° MLAT) on 8 March.

The MLATs derived from European sources show much higher value (~ 50–55°) than those from East Asian sources (~ 30–36°). There are two possible reasons. The first reason is the temporal variation of the auroral oval. The auroral oval expands equatorward during the storm main phase, and contracts during the storm recovery phase. The main phase usually lasts 1 to a few hours (Gonzalez et al. 1994). Therefore, it is highly plausible that the storm main phase took place when the East Asian sector was in darkness, and that the storm main phase ended when the European sector was in darkness. During the 1909 space weather event, the low latitude aurorae were mostly reported from Japan and Australia, due to its appearance around 12–20 UT, corresponding to the main phase of the associated magnetic storm (cf. Figure 4 of Hayakawa et al. 2019a). The second reason is the limited geographical distribution of European observers around the Mediterranean Sea. Further reports from low latitudes in the European sector, for example, North Africa, may lead to a revision of the auroral oval, but so far, we have not come across auroral reports from this region in 1582.

Figure 5 shows a drawing of the aurora at Augsburg (E07) on 6 March 1582 (cf. pp. 65–67 of Schlegel & Schlegel 2011). The drawing resembles a corona aurora that consists of a converging ray structure. The Moon is located near the focus of the corona aurora. The lunar phase is calculated as 0.40, with full moon at 0.50, on 6 March 1582, according to Kawamura et al. (2016). Despite the lunar phase of 0.40, the aurora was visible near the Moon. This means that the aurora was bright, at least on 6 March 1582.

Such a bright auroral display around the night of a full moon is confirmed by parallel auroral records such as those at Yerkes Observatory (N42°34', W088°33') on 4 September 1908, reported by Barnard (1910, p. 223); and at Ashurst (N51°16', W000°01'; 55.0° MLAT) on 18 February 1837, and Dulwich Wood (N51°26', W000°04'; 55.1° MLAT) on 12 November 1837, reported by Snow (1842, p. 9 and p. 11); as reviewed by Stephenson et al. (2019). Ebihara et al. (2017) explained the extremely bright aurora (~ IBC class IV) at low to mid MLAT as precipitation of high-intensity low-energy electrons.

Figure 6 also shows a drawing of the aurora (E10), which shows only the vertically-extended ray structure. The shape of the aurora is different from that shown in Figure 5. The difference may come from the temporal variation of the auroral oval.

**6 SCALE OF THE MAGNETIC STORM OF 6–8 MARCH 1582**

We estimate that the equatorward boundary of the auroral oval extended down to ~33.0° ILAT. Since the equatorward boundary of the auroral oval is well correlated with the Dst index (Yokoyama et al. 1998), it can be used as a proxy for the measure of the strength of magnetic storms. The equatorward boundaries estimated in this study are higher than those of 1/2 September 1859 (28.5° or 30.8° ILAT) (Hayakawa et al. 2018d) and 4/5 February 1872 (24.2° ILAT) (Hayakawa et al. 2018b), comparable with those of 25/26 September 1909 (31.6° ILAT) (Hayakawa et al. 2019a) and 28/29 August 1859 (36.5° ILAT) (Hayakawa et al. 2018d), and lower than those of 13/14 March 1989 (40° ILAT) (Allen et al. 1989). With the aid of two empirical formulae suggested by Yokoyama et al. (1998), we estimated the Dst values for ILATs of 33.0°, 34.9°, and 37.9° as follows. For formula 1, the Dst values are −754, −658, and −519 nT, respectively. For formula 2, they are −1120, −975, and −761 nT, respectively. However, care is needed when one deals with extreme magnetic storms. The formula 1 was derived based on the equatorward boundary observed by the DMSP satellite in 1983–1991 (Gussenhoven, Hardy & Burke 1981). There are only three data points for a Dst value less than −300 nT. The formula 2 was derived based on the observation on 13/14 March 1989. Therefore, the formulae may be ambiguous for extreme magnetic storms. We conservatively consider that the minimum Dst value of the magnetic storm on 8 March 1582, at least, is comparable to those on the storm on 13/14 March 1989 (−589 nT) and the storm on 25/26 September 1909 (−595 nT) (Love et al. 2019; Hayakawa et al. 2019a).

We also show that the aurora was visible at low- and mid-latitudes for three successive nights, from 6 to 8 March, on the basis of auroral observations in Europe. This suggests that the magnetic storm of March 1582 lasted at least 3 days, which is probably caused by a series of ICMEs (interplanetary coronal mass ejections) such as the magnetic storm on 13/14 March 1989. The storm on 13/14 March 1989 was probably associated with a series of solar flares including the X4.0 flare on 9 March, the X4.5 flare on 10 March, and the X6.5 flare on 17 March (Allen et al. 1989). A few days earlier, on 6 March, an X15 flare occurred, which was the second largest X-ray flare in solar cycle 22 (Watari, Kunitake & Watanabe 2001). This situation is similar to the Halloween sequence in 2003, when multiple CMEs from AR 10486 hit the terrestrial magnetic field and caused multiple magnetic storms. The initial ICME is thought to have cleaned up the interplanetary space or to have interacted with the following ICMEs to make the following ICMEs more geo-effective (Gopalswamy et al. 2005a, 2005b; Gopalswamy 2018; Lefèvre et al. 2016; Shiota & Kataoka 2016) and to have triggered 'perfect storms' by their combined effect (Lefèvre et al. 2016; Hayakawa et al. 2017d, 2018d; Liu et al. 2019). Likewise, the ICME that caused the magnetic storm on 28/29 August 1859 probably cleaned up interplanetary space, allowing the following ICMEs to travel without much deceleration, and producing the Carrington storm on 1/2 September (Hayakawa et al. 2018d). Therefore, we can reasonably infer that the preceding ICME(s) caused auroral displays on 6/7 March 1582 at mid-latitudes (in European sector) and cleaned up the interplanetary space to allow the following ICME(s) to cause the magnetic storm on 8 March 1582 with its auroral display visible down to ~33.0° ILAT, on the basis of case studies of similar space weather events.

**7 CONCLUSION**

In this paper, we examined auroral records of March 1582 in East Asia and Europe. We showed that the equatorward boundary of auroral visibility was as low as ~28.8° MLAT on 8 March 1582. The equatorward boundary of the auroral oval was located at ~33.0° ILAT. This is comparable to other known magnetic storms on 25/26 September 1909 (~31.6° ILAT, minimum Dst of −595 nT) and 28/29 August 1859 (~36.5° ILAT), and 13/14 March 1989 (~40° ILAT, minimum Dst of −589 nT). The European records show preceding auroral displays, suggesting that preceding ICMEs may play an important role in making the following ICMEs more geo-effective and producing the low-latitude aurora on 8 March 1582. The situation in interplanetary space may be similar to the Halloween storms of October 2003 and the storms of August and September 1859.

While these space weather events occurred before the beginning of the telescopic sunspot observations in 1610, we calculate that they took place 34 years – roughly 3 solar cycles – before the first cycle maximum within the coverage of the telescopic sunspot observations (e.g. Usoskin, Mursula & Kovaltsov 2003; Vaquero et al. 2011). In comparison with another space weather event in 10 September 1580 (Kázmér & Timár 2016), the 1582 events may have fallen in the declining phase following the solar maximum around 1581, as the exact length of each solar cycle may differ from the mean 11-year cycle duration. This topic is worth further considerations on the basis of the solar cycle reconstruction derived from cosmogenic isotopes, which seems to locate this event in a declining phase of the solar activity (I. G. Usoskin, private communication), as is usually the case with large space weather events (Kilpua et al. 2015; Lefèvre et al. 2016). The results provide insight into space weather events that occurred before the beginning of sunspot observations in 1610, and a basis for discussing the long-term solar activity since then.

**ACKNOWLEDGEMENT**

We thank I. G. Usoskin for his advice on the phase of the solar cycle in 1582, on the basis of the reconstruction with cosmogenic isotopes, GMT (Wessel & Smith 1991) for helping us in creating maps, and A. D. Kawamura for helping the calculation of lunar phase. The authors wish to express their sincere appreciation and acknowledgement to RISH, Kyoto University, and its financial support of mission-linked research. This study was supported by JSPS KAKENHI grants JP15H03732, JP15H05815, JP17J06954, and JP18H01254. We also thank Choryuji Temple and Hakusan Culture Museum for permission of reproduction of the photo of J02, M. Suzuki of Hakusan

Hattori, Hayakawa, and Ebihara (2019) Great Magnetic Storms in 1582
*Monthly Notices of the Royal Astronomical Society*, doi: 10.1093/mnras/stz1401

Hattori, Hayakawa, and Ebihara (2019) Great Magnetic Storms in 1582
*Monthly Notices of the Royal Astronomical Society*, doi: 10.1093/mnras/stz1401

**Table 1**. List of observing sites from East Asia. Shown are record ID, date (year, month, and date), colour, direction, observational time of start and end, geographical latitude and longitude, magnetic latitude (MLAT) and location. Each original reference is given in Appendix 1. Colour is given as R (red), W (white) and B (blue). Time is local time between 06:00 and 30:00 corresponding to 06:00 on the next day. We identified the observing sites according to the locations of their authors at that time (K01, K02, J02–J06), unless otherwise the observing sites are explicitly clarified in the title (J01) or the text (C01, C02, J07).

| ID | year | mon. | date | colour | direction | start | end | lat. | long. | MLAT | location |
|---|---|---|---|---|---|---|---|---|---|---|---|
| J01 | 1582 | 3 | 8 | R | wnw−ene | 20:00 | 24:00 | N36°09′ | E140°31′ | 31.6 | Hokota |
| J02 | 1582 | 3 | 8 | R | ne | 22:00 | 26:00 | N35°55′ | E136°50′ | 31.4 | Choryuji |
| J03 | 1582 | 3 | 8 | R | − | night | − | N35°01′ | E135°45′ | 30.5 | Kyoto |
| J04 | 1582 | 3 | 8 | R | n | night | − | N35°01′ | E135°45′ | 30.5 | Kyoto |
| J05 | 1582 | 3 | 8 | R | n−z | night | − | N34°41′ | E135°50′ | 30.2 | Kōfukuji |
| J06 | 1582 | 3 | 8 | R | ne | 20:00 | − | N34°41′ | E135°50′ | 30.2 | Kōfukuji |
| J07 | 1582 | 3 | 8 | R | e | 22:00 | dawn | N35°09′ | E136°08′ | 30.6 | Azuchi |
| J07 | 1582 | 3 | 8 | R | − | − | − | N33°14′ | E131°36′ | 28.8 | Búngo |
| K01 | 1582 | 3 | 8 | R/W/B | nw→e→s | 20:00 | dawn | N36°41′ | E128°25′ | 32.3 | Yecheon |
| K02 | 1582 | 3 | 8 | R/W/B | nw→e→s | 20:00 | dawn | N36°41′ | E128°25′ | 32.3 | Yecheon |
| C01 | 1582 | 3 | 8 | R/W | z | − | − | N38°56′ | E100°27′ | 35.6 | Gānsù |
| C02 | 1582 | 3 | 8 | R/W | − | − | − | N38°56′ | E100°27′ | 35.6 | Gānsù |

**Table 2**. List of observing sites from Europe. Shown are record ID, date (year, month, and date), colour, direction, observational time of start and end, geographical latitude and longitude, magnetic latitude (MLAT) and location. Each original reference is given in Appendix 2. Colour is given as R (red), W (white), B (blue), Y (yellow) and Gr (green). Time is local time between 06:00 and 30:00 corresponding to 06:00 on the next day. We identified observing sites according to the location of their authors at that time (E10, E16, E22, E24, E25, E27, E35 and E39), unless the observing sites are explicitly mentioned in the title or the text.

| ID | year | mon. | date | colour | direction | start | end | lat. | long. | MLAT | location |
|---|---|---|---|---|---|---|---|---|---|---|---|
| E01 | 1582 | 3 | 6 | R | − | night | − | − | − | − | Behmerland |
| E01 | 1582 | 3 | 6 | R | − | night | − | − | − | − | Schlesien |
| E02 | 1582 | 3 | 6 | R | − | night* | − | N49°40' | E014°26' | 52.5 | Sedlčany |
| E03 | 1582 | 3 | 6 | R | e→z→nw | 19:00 | − | N47°23' | E008°33' | 50.6 | Zürich |
| E04 | 1582 | 3 | 6 | R | n | − | − | N47°23' | E008°33' | 50.6 | Zürich |
| E05 | 1582 | 3 | 6 | R | e→z→nw | 19:00 | − | N47°23' | E008°33' | 50.6 | Zürich |
| E06 | 1582 | 3 | 6 | R | − | 19:00 | − | N51°10' | E013°30' | 54.1 | Meissen |
| E07 | 1582 | 3 | 6 | R/W/Y | e−z | 21:00 | 24:00 | N48°22' | E010°54' | 51.5 | Augsburg |
| E08 | 1582 | 3 | 6 | R/W | − | 24:00 | − | N46°57' | E007°27' | 50.3 | Bern |
| E09 | 1582 | 3 | 6 | R | − | 18:30 | − | N50°43' | E012°28' | 53.7 | Zwickau |
| E10 | 1582 | 3 | 6 | R/W/Y | e→z | 19:00 | 23:00 | N51°00' | E013°27' | 53.9 | Mohorn |
| E11 | 1582 | 3 | 6 | R | − | − | − | N50°14' | E012°52' | 53.2 | Karlsbad |
| E12 | 1582 | 3 | 6 | R | − | night* | − | N44°04' | E012°34' | 47.1 | Rimini |
| E13 | 1582 | 3 | 6 | R/W | w−n | 19:00 | − | N46°07' | E004°57' | 49.6 | Châtillon-les-Dombes |
| E14 | 1582 | 3 | 6 | R | z | 19:00 | − | − | − | − | Bourgogne, Bresse etc. |
| E15 | 1582 | 3 | 6 | R | − | night* | 20:00 | N45°08' | E008°27' | 48.4 | Casale Monferrato |
| E16 | 1582 | 3 | 6 | various | − | − | − | N47°13' | E007°32' | 50.5 | Solothurn |
| E17 | 1582 | 3 | 6 | − | − | 19:00 | − | N47°44' | E010°19' | 50.9 | Kempten |
| E18 | 1582 | 3 | 6 | R | z | night* | − | N50°19' | E013°30' | 53.3 | Žatec |
| E19 | 1582 | 3 | 6 | R | z | evening | night | N50°05' | E010°34' | 53.2 | Königsperg |
| E19 | 1582 | 3 | 6 | R | z | evening | night | N47°45' | E007°20' | 51.1 | Mülhausen |
| E20 | 1582 | 3 | 6 | R | − | 20:00 | − | N51°20' | E012°22' | 54.3 | Leipzig |
| E21 | 1582 | 3 | 6 | R | − | 20:00 | 23:00 | N46°16' | E006°03' | 49.7 | Genève |
| E22 | 1582 | 3 | 6 | R/W | − | 19:00 | 22:00 | N51°03' | E013°44' | 54.0 | Dresden |
| E23 | 1582 | 3 | 6 | R | − | 20:00 | − | N47°59' | E010°11' | 51.1 | Memmingen |
| E24 | 1582 | 3 | 6 | R/W | − | 20:00 | 21:00 | − | − | − | Budtstadt |
| E25 | 1582 | 3 | 6 | R | − | night* | − | N47°13' | E007°32' | 50.5 | Solothurn |
| E26 | 1582 | 3 | 6 | R | − | 20:00 | − | N44°56' | E004°54' | 48.4 | Vallence |
| E27 | 1582 | 3 | 6 | R | n | night | − | N45°11' | E005°13' | 48.6 | Saint-Antoine |
| E28 | 1582 | 3 | 6 | R | n | 19:00 | − | N46°42' | E006°55' | 50.0 | Romont |
| E29 | 1582 | 3 | 6 | R | − | 20:00 | − | N40°25' | W003°42' | 44.3 | Madrid |
| E30 | 1582 | 3 | 6 | R | − | night | − | N40°25' | W003°42' | 44.3 | Madrid |
| E21 | 1582 | 3 | 7 | R | − | 20:00 | 23:00 | N46°16' | E006°03' | 49.7 | Genève |
| E31 | 1582 | 3 | 7 | R/W | − | 19:00 | 26:00 | N46°59' | E006°53' | 50.3 | Auvernier |
| E32 | 1582 | 3 | 7 | R | − | (2-hour duration) | | N43°17' | E011°59' | 46.3 | Cortona |
| E33 | 1582 | 3 | 7 | R | − | − | − | − | − | − | Braband, Holand etc. |
| E05 | 1582 | 3 | 8 | − | − | night | − | N47°23' | E008°33' | 50.6 | Zürich |
| E19 | 1582 | 3 | 8 | R | − | evening | − | − | − | − | Preussen |
| E34 | 1582 | 3 | 8 | R | − | − | − | N51°32' | E009°55' | 54.7 | Göttingen |
| E35 | 1582 | 3 | 8 | R/Gr/B etc. | − | 20:00 | 22:00 | N48°34' | E003°18' | 52.1 | Provins |
| E36 | 1582 | 3 | 8 | R | − | 21:00 | − | N48°51' | E002°21' | 52.4 | Paris |
| E37 | 1582 | 3 | 8 | R | w | 21:00 | − | N47°21' | E000°51' | 51.0 | Azay-sur-Cher |
| E38 | 1582 | 3 | 8 | various | − | 21:00 | − | N51°13' | E004°24' | 51.3 | Anvers |
| E39 | 1582 | 3 | 8 | R | n−w | 21:00 | − | N51°30' | E000°00' | 55.2 | London |

*The original descriptions of time in E02, E12, E15, E18 and E25 are 'po prwní hodině na noc', 'un' hora di notte', 'un'ora di notte', 'initio nocte' and 'sub primam noctis', respectively.

**Figure 1.** Auroral descriptions of East Asia records on 8 March 1582. (a) Reprint of the original record of J02 in Choryuji Temple (N35°55', E136°50', 31.4° MLAT). Choryuji Temple owns the original record and Hakusan Culture Museum owns its photo. We reproduced the photo by permission of Choryuji Temple and Hakusan Culture Museum. (b) Reprint of K01 in Yechon (N36°41′, E128°25′, 32.3° MLAT). [available only in the record version]

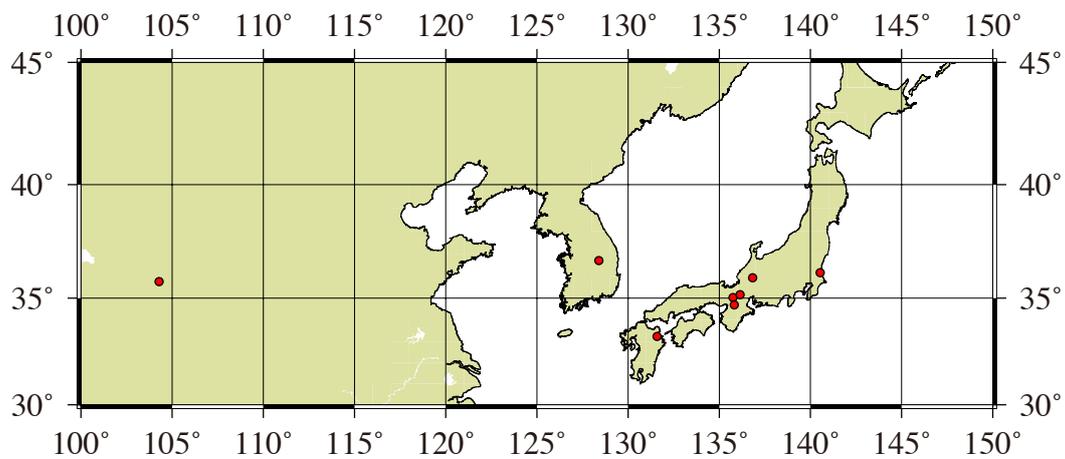

**Figure 2**. Observing sites of auroral displays on 8 March 1582 in East Asia. The solid lines show geographic latitude and longitude.

Hattori, Hayakawa, and Ebihara (2019) Great Magnetic Storms in 1582
*Monthly Notices of the Royal Astronomical Society*, doi: 10.1093/mnras/stz1401

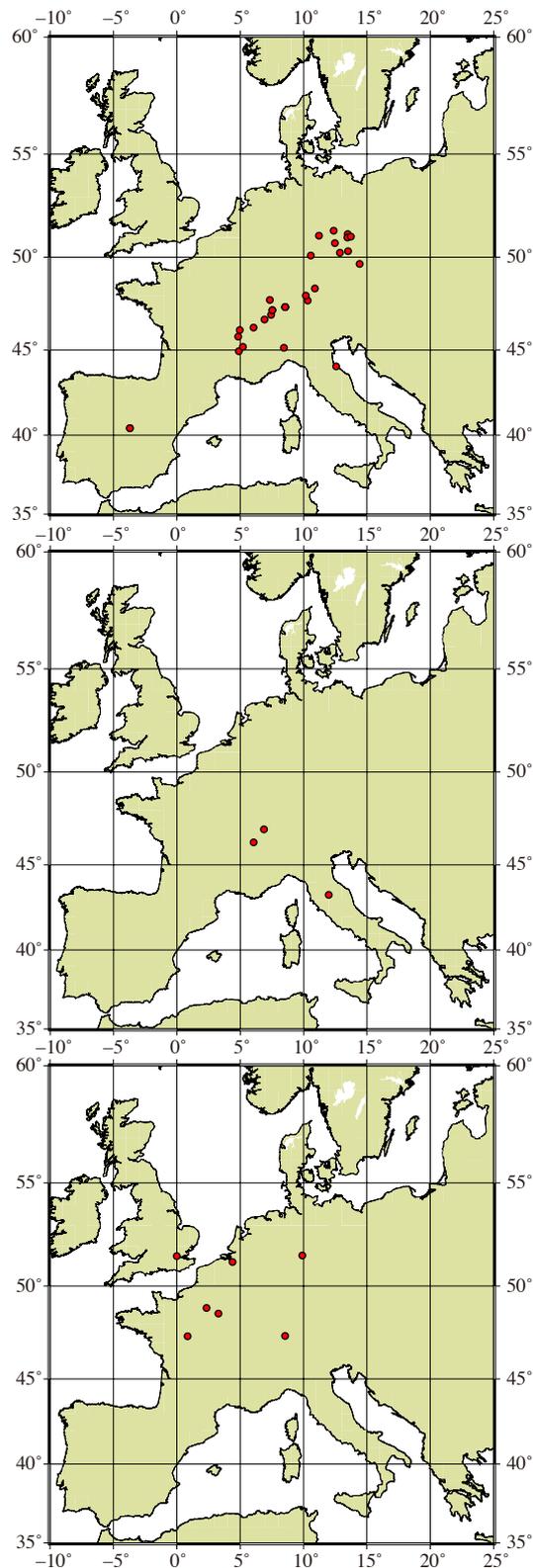

**Figure 3**. Observing sites of auroral displays during 6 (above), 7 (middle), and 8 (below) March 1582 in Europe. The solid lines show geographic latitude and longitude.

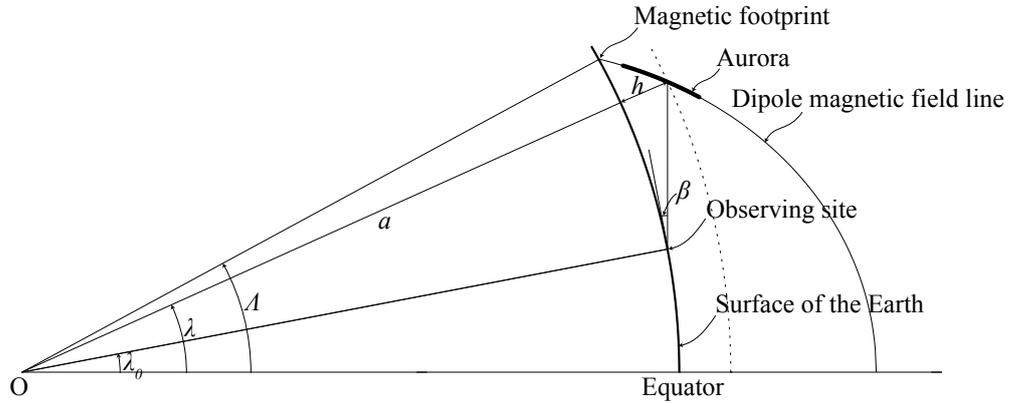

**Figure 4**. Relationship between the magnetic latitude (MLAT) of an observing site $\lambda_0$, and the invariant latitude (ILAT) of the aurora $\Lambda$ at the height $h$ in dipole geometry, modified from figure 2 of Hayakawa et al. (2018d). The elevation angle of the auroral display is $\beta$. $R$ is the geocentric distance of the aurora, which is a sum of the radius of the earth $a$ and the height of the aurora $h$.

**Figure 5**. Auroral drawing of E07 in Augsburg (N48°22', E010°54', 51.5° MLAT) on 6 March 1582. The original figure (a single-sheet woodcut) is reproduced by permission of Zentralbibliothek Zürich, Department of Prints and Drawings/ Photo Archive (shelf mark: ZB Graphische Sammlung (GSB), PAS II 19/4). [available only in the record version]

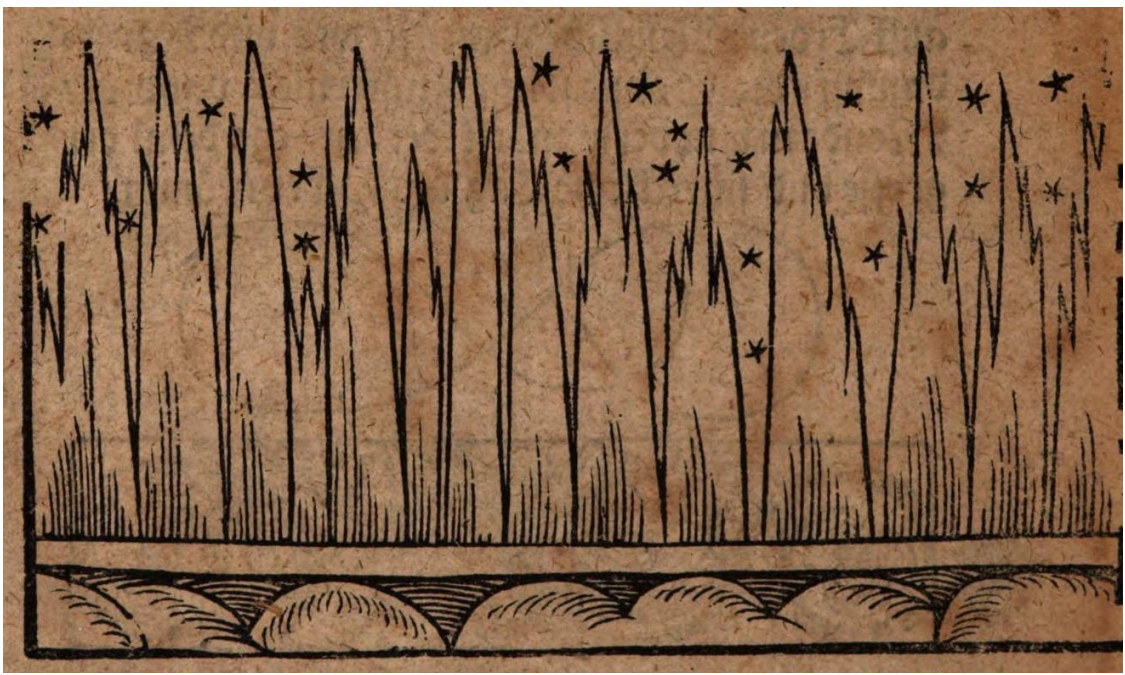

**Figure 6**. Auroral drawing shown below the title of E10 on 6 March 1582. The original figure is reproduced from the digital collections of Bayerische StaatsBibliothek (shelf mark: Res/Astr.p. 207 u).

**APPENDIX**

Historical sources in Appendix 1 are in Chinese (C), Korean (K) and Japanese or Portuguese (J), while those in Appendix 2 are in European languages. For records of Appendix 1 except for J07, we provided the title, page (or folio) number of auroral descriptions used for the table, name of publication, and year of publication. For J07 and records of Appendix 2, we basically provide the author, year of publication, title, journal (if any) and page (or folio) number of auroral descriptions used for the table. For manuscripts (E05 E22, E25 and E29), we provide the title and folio (or page) number of auroral descriptions used for the table. We also added the location and shelf mark surrounded by parentheses to the end of manuscripts and contemporary publications (E07, E09, E10, E13, E14 and E33).

**APPENDIX 1: REFERENCES OF HISTORICAL SOURCES FOR EAST ASIAN AURORAL OBSERVATIONS**

– C01: 神宗顯皇帝實錄, v. 121, p. 2281 (v. 131, p. 2447); 明實錄 ／ 中央研究院歷史語言研究所校印 v. 101 (v. 102), 1966

– C02: 國榷, v. 71, p. 4404; 國榷 第五冊, 1958

– K01: 草澗日記, p. 572; 韓國學資料叢書 v. 14, 1997

– K02: 草澗先生文集卷之四 雜記, p. 278; 韓國歷代文集叢書 v. 97, 1993

– J01: 烟田旧記, p. 285; 鉾田町史 中世史料編 烟田氏史料, 1999

– J02: 荘厳講執事帳, v. 2, p. 249; 白鳥町史 史料編, 1973

– J03: 晴豊記, f. 23r; 文科大学史誌叢書 晴豊記 v. 1, 1899

– J04: 立入左京亮入道隆佐記, p. 286; 史籍集覽 v. 13, 1902

– J05: 蓮成院記録, p. 250; 多聞院日記 v. 5, 1939

– J06: 多聞院日記, p. 203; 多聞院日記 v. 3, 1936

– J07: Manuel de Lyra, 1598, *Cartas que os padres e irmãos da Companhia de Jesus, escreuerão reynos de Japão & China rão dos reynos de Japão & China aos da melma companhia da India, & Europa, des do anno de 1549 atè o de 1580*, v. 2, f. 63r

**APPENDIX 2: REFERENCES OF HISTORICAL SOURCES FOR EUROPEAN AURORAL OBSERVATIONS**

– E01: Schlesinger L., 1881, *Simon Hüttels Chronik der Stadt Trautenau (1484-1601)*, p. 259

– E02: Březan V., 1847, *Wácslawa Březana Život Wiléma z Rosenberka*, p. 251

– E03: Scheuchzer J. J., 1746, *Natur-Geschichte des Schweitzerlandes*, v. 1, p. 75

– E04: Wolf R., 1856, Ergänzungen zu Mairan's Liste des apparitions de l'Aurore boréale, *Vierteljahrsschrift der Naturforschenden Gesellschaft in Zürich*, v. 1, p. 196

– E05: *Sammlung von Nachrichten zur Zeitgeschichte aus den Jahren 1560-87*, v. 20, ff. 169r–169v (Zentralbibliothek Zürich, Ms F 30) doi:10.7891/e-manuscripta-17668

– E06: Lehmanns C., 1699, *Christian Lehmanns Sen. weiland Pastoris zu Scheibenberg Historischer Schauplatz derer natürlichen Merckwürdigkeiten in dem Meißnischen Ober-Ertzgebirge…*, p. 418

– E07: Schultes H., 1582, *Warhafftige und erschröckliche Newe Zeyttung so sich am Himmel erzeyget hat den 6. Martij Anno 1582 Jar vngefahrlich von 9. biß auff 12. uhr gestanden*, f. 1r (Zentralbibliothek Zürich, Department of Prints and Drawings/ Photo Archive, ZB Graphische Sammlung (GSB), PAS II 19/4)

– E08: Wolf R., 1857, Auszug aus dem Chronicon Bernensi Abrahami Musculi ab Anno 1581 ad Annum 1587, *Mitteilungen der Naturforschenden Gesellschaft in Bern*, p. 108

– E09: Moller T., 1582, *Kurtze Beschreibung des erschrecklichen Zeichens, so erschienen im Mertzen, Dieses Jars*, f. 1r (Österreichische Nationalbibliothek, 39.R.30)

– E10: Bapst M., 1582, *Warhafftige beschreibung Des Erschrecklichen Blut vnd Fewerzeichens, welches den 6. Martij dieses ablauffenden 1582. Jahrs am Himmel gesehen worden…*, ff. 1r–1v (Bayerische StaatsBibliothek, Res/Astr.p. 207 u)

– E11: Weller E., 1872, *Ersten Deutschen Zeitungen herausgegeben mit einer bibliographie (1505-1599)*, p. 272

– E12: Tonini L., 1887, *Rimini dal 1500 al 1800*, p. 346

– E13: De La Tayssonnière G., 1582, *Le grand horrible et espouventable meteore apparu au ciel le mardy sixiesme jour du present mois de mars 1582*, pp. 6–10

– E14: Morel C., 1582, *Vision, et signes prodigieux, apparuz & veuz és Conté, & Duché de Bourgõgne, pays de Bresse, & lieux circonvoisins, le sixiesme jour du moys de Mars mil cinq cens quatre vingts deux*, pp. 6–9 (Bibliothèque de l'Arsenal, 8°-S-1506)